\documentclass[prb,twocolumn,showkeys,showpacs,preprintnumbers,amssymb,amsmath]{revtex4}

\usepackage{graphicx}
\usepackage{bm}
\usepackage{dcolumn}
\begin{document}

\title{Theoretical study of dislocation nucleation from simple surface defects
in semiconductors} 

\date{\today}

\author{J. Godet}

\email{julien.godet@etu.univ-poitiers.fr}

\author{L. Pizzagalli}

\author{S. Brochard}

\author{P. Beauchamp}

\affiliation{Laboratoire de M\'etallurgie Physique, CNRS UMR 6630, Universit\'e 
de Poitiers,  B.P. 30179, 86962 Futuroscope Chasseneuil Cedex, France}

\begin{abstract}
dislocation nucleation
\end{abstract}

\pacs{pacs codes}

\keywords{dislocation; nucleation; silicon; simulation}

\preprint{report number}

\maketitle

\section{Introduction}

The plasticity of semiconductors has been a subject of numerous studies for the
last decades  in both fundamental and applied research. Despite significant
progress in the  understanding of the fundamental mechanisms involved, several
issues remain, in particular for  nanostructured semiconductors. In these
materials, including for example  nano-grained systems or nanolayers in
heteroepitaxy,  dimensions are usually too small to allow the classical
mechanisms of  dislocation multiplication, such as Franck-Read
sources.\cite{Mou99PMA} It is then likely  that other mechanisms dominate, and
it has been already proposed that surfaces  and interfaces, which become
prominent for small dimensions, play a major role.  Several observations
support this assumption, especially for strained layers and misfit
dislocations  at interfaces.\cite{Dun97JMSME,Jai97PMA,Wu01PMA} The question of
dislocation formation at surfaces concerns also bulk materials submitted to
large stresses. \cite{Wu98PML,Sca99PSS,Rab01MSE}   The propagation of
dislocation from surfaces  has been investigated in the frame of a continuum
model and elasticity theory.  However, the characterization of the nucleation
of dislocations is out of the reach of this  approach, and the predicted
activation energy is very large, in disagreement with experiments.  It is also
difficult to investigate experimentally the very first stages of dislocation
formation.  Hence, the mechanisms involved in the nucleation of dislocations
from surfaces or interfaces are far to  be well understood. There has been some
attempts to perform atomistic calculations for addressing this  issue.  In
particular, the interaction between a dislocation and the free surface or the
interface,\cite{Ich95MSE,Asl98PML} or between  ledges and a crack tip,
\cite{Jua96PML} and  the instability of a stressed ledge \cite{Gao99PMA} have
been studied.

It has been proposed that surface defects like steps, or cleavage ledges, could
favor the nucleation  of dislocations, by lowering the activation
energy.\cite{Xu97PMA} This assumption is supported by experimental facts,  with
dislocation sources located on the cleavage surface and coinciding with
cleavage ledges.  \cite{Gal01PMA,Arg01SM2} Atomistic simulations of dislocation
nucleation from surface defects  in metals have also been recently reported.
\cite{Bro00PMA} It has been shown that the presence of the step  modifies the
otherwise uniform strain field,\cite{Bro00PRB} which effectively makes easier
the dislocation formation. The situation appears to  be different for
semiconductors, with no clear strain inhomogeneity at the step.
\cite{Poo92PRB,God02SM2} The role  of the stress orientation on the dislocation
formation is also unclear. Additional atomistic simulations  are needed to shed
light on these points and fully characterize the mechanisms behind dislocation
nucleation. 

In this paper, we report large-scale atomistic calculations of the nucleation
of dislocations from surface  defects in systems submitted to a stress with
variable orientation.  We focused on linear surface defects, with simple steps
but also  cleavage ledges. As for the material, silicon was selected as the
best candidate, for several reasons. First,  it is a good model, since a lot of
semiconductors crystallize in the same cubic diamond structure, or  the
zinc-blende structure, almost equivalent from the point of view of plasticity. 
Second, silicon can be grown without dislocations, which allows  a comparison
between experiments and simulations. Finally, several high quality atomistic
potentials are  available. In the first part of the paper, the silicon
structure and the slip systems are briefly described. After the presentation of
the model and the calculations techniques used to perform the simulations, the
results obtained with several empirical potentials are described. In
particular, we mostly focus on  stress orientations that increase the
probability of nucleating the relevant dislocations. Several points are then
discussed, such as the conditions of nucleation, the role of stress orientation
and  temperature, and the slip system selected. 

\section{Methodology}

\subsection{Structure and geometry}

In ambient conditions, the stable structure of silicon is diamond cubic.
Dislocations glide in the (111) dense planes, that are gathered together in two
sets, one widely spaced, called "shuffle" set and one narrowly spaced, called
"glide" set (Fig.~\ref{slip_sys}a). The Burgers vector of a perfect dislocation
is $1/2<1\bar{1}0>$. Dissociation only occurs in the glide set, with two
Shockley partial dislocations, $1/6<1\bar{2}1>$ and  $1/6<2\bar{1}\bar{1}>$.
\cite{Hir82WIL} Note that, since the Burgers vector of a partial dislocation
doesn't join two points of the crystal lattice, the nucleation of a partial
dislocation is always accompanied by a stacking fault of the atomic plane. When
occurring on adjacent atomic planes, the stacking faults form micro-twins.
Considering the angle between the dislocation line and the Burgers vector,
perfect dislocation are called $60^\circ$ or screw, while partials are called
$90^\circ$ or $30^\circ$.  These notations are used in the following.

A semi infinite system including surface steps is modeled by employing a slab
with a $2\times1$ rebuilt (100) free surface (Fig.~\ref{cell}).
\cite{Cha79PRL,Ram95PRB} Four atomic layers are frozen in the bottom of the
slab, opposite to the free surface. Steps, lying along the $[0\bar{1}1]$ dense
directions, which correspond to the intersection of \{111\} slip planes and the
(100) surface, are placed on the free surface. The steps are made infinite
through the use of periodic boundary conditions along the $[011]$ direction.
Two steps of opposite signs are introduced in order to allow the use of
periodic boundary conditions in the $[0\bar{1}1]$ direction normal to the step
line.  The dimensions along $[100]$ and $[011]$ have been determined by 
several calculations on systems  with different sizes, for minimizing the
interactions between the surface and the frozen bottom, and between steps
(Fig.~\ref{cell}). A typical system encompasses  4 atomic layers along the step
line direction $[0\bar{1}1]$, 120 along the surface normal $[100]$ and 160
along $[011]$, the normal to the step in the surface plane, i.e. about 80000
atoms.

Note that the periodicity of 4 atomic layers along the step direction is a
severe limitation of the simulation in that it almost restricts the problem to
two dimensions, and prevents in particular the formation and expansion of such
defects as dislocation half-loops.

In this work, the most simple steps formed by the emergence of a perfect
dislocation at the surface are considered. They are called $D_{B}$ re-bonded
and $D_{B}$ non re-bonded,\cite{Cha87PRL} and have a height of two atomic
layers. The effect of higher steps is also checked by considering cleavage
ledges corresponding to 5 $D_{B}$ step forming a \{111\} facet.

\subsection{Application of a uniaxial stress}

To simulate the effect of an applied uniaxial stress $ \bm{\sigma}$, the system
is deformed with strains calculated using the silicon compliances $S_{ijkl}$.
The latter are obtained from the elastic constants $C_{ijkl}$, computed for all
empirical potentials. In this work, the uniaxial stress direction is contained
into the surface, but its orientation with respect to the step line, can vary.
As a result, the projection of this stress  in the \{111\} slip planes, called
the resolved shear stress, will also vary. This quantity is important since it
is reasonable to assume that the slip system with the largest resolved shear
stress along the Burgers vector {\bf b} will be favored. The relationship
between the resolved  shear stress $\bm{\tau}$ and the uniaxial stress
$\bm{\sigma}$ is $|\bm{\tau}| = \pm s|\bm{\sigma}|$, $s = cos \varphi\ cos \nu$
being the Schmid factor. $\varphi$ is the angle between $\bm{\sigma}$    and
the normal of the slip plane and $\nu$ the one between $\bm{\sigma}$ and {\bf
b}. In the Fig.~\ref{Schmid}, the calculated Schmid factors along several slip
directions into the \{111\} slip planes are represented as a function of the
angle $\alpha$ between the stress orientation and $[011]$, the normal to the
step lines. Four dislocations are possible: the $60^\circ$ and screw perfect
dislocations, and the $90^\circ$ and $30^\circ$ partials. The most efficient
stress orientations for each dislocation are gathered in the
Fig.~\ref{slip_sys}b. The maximum resolved shear stress along the Burgers
vector of the $60^\circ$ (screw) is obtained for $\alpha=22.5^\circ (45^\circ)$
for both tensile and compressive stress, respectively. The $90^\circ$ is
favored in case of a non disorientated tensile stress only. A compressive
stress would give a resolved shear stress in the anti-twinning sense. Finally,
the $30^\circ$ is favored by a disorientated stress of $36^\circ$, only in
compression to produce a twinning stress. 

\subsection{Computational methods}

The large number of atoms required in the simulation prevents the use of ab
initio methods because it would be too expensive in CPU time. Instead, three
classical potentials for silicon are employed: the potential  of Stillinger and
Weber (SW),\cite{Sti85PRB} based on a linear combination  of two- and
three-body terms, the Tersoff potential \cite{Ter89PRB} including many-body
interactions thanks to a bond order term in the functional form, and the
environment-dependent inter-atomic potential \cite{Baz97PRB} (EDIP), more
recent and designed specifically for simulating defects. 

To deform the system, stress increments of 1.5 GPa (equivalent to a strain
around 1 to 1.4\% according to the stress orientation) are successively
applied, the atomic positions being relaxed between each increment. Two
relaxation techniques are used. Either, a static relaxation with a conjugate
gradients algorithm is performed, until forces on atoms are smaller than
$10^{-3}$ eV/\AA, or temperature is introduced in simulations \cite{Rif99XMD} 
with molecular dynamics, in order to investigate its effect on the nucleation.
After an initial static relaxation with conjugate gradients, temperature is
introduced by increment of 300K, with a simulation time ranging from 5 to 50
ps. 

\section{Results with the Stillinger-Weber potential}

Three temperature domains have been considered; the first one at 0K, the second
one for low temperatures ($\alt$ 900K) and the  last one for high temperatures
($\agt$ 900K). In each case, we focused on relevant stress orientations, in
particular those that increase the probability of nucleating the four possible
dislocations (60$^\circ$, screw, 90$^\circ$ and 30$^\circ$). Few other stress
orientations have also been checked. All results are summarized in the
Table~\ref{SW_results}. All the cases presented here  concern  systems with
D$_B$ non re-bonded surface steps. The main effect of higher steps, like
cleavage ledges, is a slight decrease of the elastic limits. In addition, the
plastic events remain qualitatively similar. 

\subsection{Calculations at 0 K; effect of stress orientation}

At 0K, the plastic events appear under large strains in both compression and
traction, i.e. greater than 7\% (10.5 GPA) (Table~\ref{SW_results}). They are
initiated from the surface, in the close neighborhood of the step. Note that
the elastic limits for stressed systems with surface steps are always smaller
than for systems without surface steps. Hence the steps help for forming
plastic events, by lowering the required stress, and by confining the starting
surface area. Before the occurrence of plastic events, the system is
elastically deformed and the resulting shear strains are mainly located in
shuffle set planes. 

Investigations have been performed with stress orientations favoring the
nucleation of perfect dislocations. For the 60$^\circ$ dislocation, the most
efficient orientation is $\alpha = 22.5^\circ$ both in traction and in
compression (Fig.~\ref{slip_sys}b). The results in traction show a relatively
large elastic limit of 22.5 GPa (18.7\%). Beyond this stress, plasticity
occurred and the relaxed system is displayed in the Fig.~\ref{sw_22.5_tract}. 
The insert at the top of the figure clearly shows that the surface step is now
twice higher. Moreover the displacements in the shuffle set plane crossing the
step correspond to the slip of a $60^\circ$ dislocation. On the second insert
into the Fig.~\ref{sw_22.5_tract}, one can see the dislocation that has stopped
on the bottom of the simulation box and another $60^\circ$ dislocation with the
same screw component occurring in the  symmetric \{111\} shuffle set plane from
the frozen bottom. Since the dislocation is stopped on the frozen zone (which
mimics the bulk) the system must find another slip system to continue its
relaxation. In compression, large plastic strains appear from the surface steps
for a strain of around -10\% (-12 GPa),  following approximately the \{111\}
planes, but without any clearly identifiable dislocations.

The perfect dislocation in the screw orientation, should be favored by a stress
disorientated at around $45^\circ$ in both compression and traction
(Fig.~\ref{slip_sys}b). However, under a compressive stress, a 60$^\circ$
dislocation instead is nucleated in the  shuffle set plane crossing the step.
The dislocation decreases the step height and glides in the plane of the
shuffle set up to the frozen bottom of the simulation box. Under a tensile
stress, defects identified as micro-twins are formed from the surface step. It
seems that these defects are due to a peculiar behavior of the SW potential
when the resolved shear stress in the \{111\} planes is along the anti-twinning
direction.   A previous analysis has shown that these twins are formed by 
glides in two shuffle set planes with a rotation of trimers in the glide set
plane.\cite{God03UNP} In Brief, in both cases traction and compression, no
screw dislocation has been nucleated.

Then, to nucleate partial dislocations, calculations with the most efficient
stress orientations are performed. When a non disorientated tensile stress
favoring the 90$^\circ$ partial is applied on the system
(Fig.~\ref{slip_sys}b), the relaxation of the atomic positions leads to the
crystal fracture. The crack is formed from the surface step for a stress of
31.5 GPa (22.9\% ). 

The 30$^\circ$ partial dislocation is privileged by a compressive stress with
an angle of 36$^\circ$. Instead, a perfect 60$^\circ$ dislocation is nucleated
in the plane of the shuffle set crossing the surface step.  Finally, although
the stress orientations are ideal to form partial dislocation according to the
Schmid factor, none is nucleated. 

We have also checked several other configurations. In particular,  stress
orientations favoring anti-twinning configurations, one for $\alpha =  0^\circ
$ in compression and another for $\alpha = 36^\circ$ in traction.  It appeared
that for both cases, micro-twins are nucleated  from the surface steps, what
may be attributed to a somewhat odd behavior of the SW  potential. In some
cases, for a tensile stress and $\alpha=36^\circ$, we obtained  peculiar glide
events after deformation. In particular, considering a ledge and not  a single
step, the structure examination after relaxation revealed the presence  of a
60$^\circ$ and a screw dislocations. We have also investigated a system  under
tension and a disorientation  angle $\alpha=10^\circ$, for which  the resolved
shear stresses on the 90$^\circ$ and the 60$^\circ$ dislocation are the same
(Fig.~\ref{Schmid}). The result is equivalent to the situation of a
non-disorientated tensile stress, with  the fracture of the crystal. 

So it appears that at 0K, in spite of the many stress  orientations tested,
only perfect dislocations, especially 60$^\circ$, located in the  shuffle set
plane passing through the surface step are nucleated. No dislocations  in the
glide set planes have been obtained. 

\subsection{Other temperatures}

The same stress orientations have also been studied in the low temperature
domain. The main difference with the 0K study is the lowering of the elastic
limit as the temperature increases in both traction and compression. However,
the results remain qualitatively similar to what has been found at 0K. Only
perfect 60$^\circ$ dislocations are routinely nucleated. And no dislocation has
been formed in the glide set plane. Nevertheless, few differences have to be
noted. Under a compressive stress favoring  the 60$^\circ$ dislocation, i.e. at
$\alpha=22.5^\circ$, the ill-defined plastic strains obtained at 0K are
replaced by a 60$^\circ$ dislocation nucleated in the shuffle set plane. In
another case, for a stress orientation leading to a resolved shear stress in
the anti-twinning direction, i.e. at $\alpha=36^\circ$ in traction, the
simultaneous formation of the 60$^\circ$ and screw dislocation is replaced by
large strained zones near the surface step. These deformations look like a
local phase change. The last difference is obtained with a tensile stress
disorientated at about 10$^\circ$ for which the resolved shear stresses on the
90$^\circ$ and 60$^\circ$  dislocations are the same. Our results show the
nucleation of a 60$^\circ$ dislocation in the shuffle set plane crossing the
step. The dislocation glides a distance around 15 \AA\ before leading to the
fracture of the crystal.

\ 

For the high temperature domain, the elastic limits continue to decrease as
the temperature is raised. No dislocation in the planes of the glide set is
observed. However the stress in the system is now relaxed in a new manner.
Previously, at low temperature, the glide events were relatively frequent in
the plane of the shuffle set. Now at high temperatures, the glide events  in
the shuffle set planes become more and more rare as the temperature increases,
until they totally  disappear. Instead, they are replaced by disorder in the
surface along the step line looking like amorphization zones and often close to
the steps. 

\section{Results with the Tersoff potential and EDIP}

The results obtained with the SW potential have shown that only perfect
$60^\circ$ dislocations are nucleated in the shuffle set plane, and at low
temperature. A previous study on bulk silicon has shown that the Tersoff
potential and EDIP are less reliable than SW in the case of large
shear.\cite{God03JPCM}  We have restricted the investigations using these
potentials  to the stress orientations favoring the nucleation of a 60$^\circ$
dislocation, i.e. with a tensile or compressive stress at $\alpha =
22.5^\circ$. 

\ 

The calculations done with the Tersoff potential at 0K give very large elastic
limits. They are around 46.7\% (51 GPa) and -38.5\% (-42 GPa) under tensile
and  compressive  stress, respectively. In traction, the crystal periodicity
along the step line direction is lost due to large strains of the bulk looking
like the beginning of a phase transition (Fig.~\ref{ters_edip}a), leading
sometimes to a crystal crack from the surface near the step. In compression, up
to -22\%, the strains remained homogeneous. Then slight undulations appeared on
the surface up to -37\%. Finally, a plastic strain occurred  in the (011)
planes close to the surface step (Fig.~\ref{ters_edip}b). In all cases no glide
events are observed. 

Calculations have been performed at different temperatures and several applied
stresses. The only effect is the decrease of the elastic limits and the
expansion of plastic strains. However, using high steps (cleavage ledges), a
large compressive strain (-11\%) and very high temperatures ranging from 1200K
to 1500K, we managed to nucleate $60^\circ$ dislocations in the shuffle set
plane passing through the step edge (Fig.~\ref{disloc_temp}b). 

\ 

The calculations performed with EDIP at 0K also show much larger elastic
limits  than the ones obtained with SW. They are around 34.5\% (52.5 GPa)  in
traction and -8.9\% (-13.5 GPa) in compression. Under tensile stresses, a
crystal crack occurred, while under compressive stresses, the \{111\} shuffle
set plane passing through the step edge is largely sheared
(Fig.~\ref{ters_edip}c-d). This shear propagates from the surface to the slab
bottom without dislocation. When the applied strain increased neighboring
shuffle set planes are also sheared.

\section{Discussions}

\subsection{Dependency on the potentials} 

Although the same stress orientations have been checked, at 0K and non zero
temperature,  the results are often different from one potential to another. 
In order to establish which potential represents best sheared silicon, we have
recently compared these three potentials  with ab initio methods.
\cite{God03JPCM}   A homogeneous shear is imposed on \{111\} planes in a
$<110>$ direction, the amplitude of shear goes up to 122\% where the diamond
cubic structure is recovered. At each shear value, the system is relaxed in
order for the simple FCC two sublattices forming the diamond structure to reach
their relative equilibrium position. In this way, one sees how the imposed
shear is distributed in the glide set and the shuffle set respectively. When
the full amplitude of the imposed shear has been applied, the crystal structure
returns to perfect diamond cubic with the SW potential and EDIP, as well as in
the ab initio calculation, through a bond breaking and new bond formation
across the shuffle plane. However, such a bond switching is not observed with
the Tersoff potential which in these conditions, does not appear suitable for
describing dislocation nucleation.

When comparing the energy curves of bulk silicon as a function  of the
homogeneous shear strain. Only the curve of the SW potential is relatively
smooth with a shape and amplitude similar to the one calculated in DFT-LDA. The
Tersoff curve is discontinuous and the EDIP curve exhibits an angular point.
Thus, only SW can account for the atomic surrounding without energy
discontinuity when the crystal is largely strained. This feature is even more
marked when looking at derivative quantities, related to stresses. In addition,
critical values such as the theoretical shear strength are overestimated by a
factor of about two with the Tersoff potential and EDIP compared to the DFT
calculation, whereas Stillinger-Weber is much closer to ab initio.

Concerning EDIP, it has not been possible to use this potential at the large
strains considered here because of an accident occurring in the curve energy
versus shear strain which produces a shear instability of the crystal. The SW
potential is not exempt of drawbacks. When the crystal is sheared in the
anti-twinning direction, \cite{God03UNP} twinning is produced through shearing
in the shuffle set planes. We do not think that this prevents the use of the SW
potential for the other stress orientations.

Hopefully, there are indications that these inadequacies of the potentials may
become less important at high temperatures, where  dislocations can be formed
at lower imposed strains. For example, under a compressive stress with
$\alpha=22.5^\circ$,  the  twin-like defect created by the SW artefact, is
replaced by a $60^\circ$ dislocation at a smaller strain. Another example is
given by the Tersoff potential which at high temperature and for large step
height can lead to the formation of a $60^\circ$ dislocation. Temperature may
rub out unphysical irregularities in the potentials.

\subsection{Role of the surface step}

Here, we focus on the results obtained with the SW potential. Plasticity occurs
for very large strains, somewhat smaller in compression than in tension.
Although the particular crystal structure and potential may be important, one
must consider that at the very large stresses considered here,the solid may
undergo some buckling instability of course in compression and not in tention,
instability which helps dislocation formation. Others studies  are in progress
to clarify this point.

For bulk silicon, The theoretical strengths obtained with The Stillinger-Weber
potential are large, in agreement with ab initio calculations Roundy and M.L.
Cohen. \cite{Rou01PRB}  Indeed, the limits of elasticity of the systems with a
free surface presenting steps are definitely  smaller than without step. For
example at 0K, for a non disorientated stress, in tension  (in compression) the
yield strain is about 22.9\% (-7.6\%) with surface steps and about 28.3\%
(-11\%) without surface step respectively. Generally the plastic strains, such
as fracture, the glide events or the amorphization zones..., occur from the
steps or from their immediate neighborhood. In fact the presence of the step
breaks the symmetry of the system leading to some stress localization near the
step. Thus the surface step is a privileged site for plasticity.

\subsection{Slip system: glide or shuffle}

Now, we discussed whether the dislocation nucleation occurs in the glide or in
the shuffle set planes, using the results obtained with the SW potential.

In principle, the perfect $60^\circ$ and screw dislocations can be formed in
either the glide or the shuffle plane, but in our results, the dislocations are
nucleated only in the planes of the shuffle set. The simulation with the stress
orientation most appropriate for nucleating $90^\circ$ and  $30^\circ$ partials
in the glide set, lead to the fracture of the crystal and to the formation of
a  $60^\circ$ dislocation in the shuffle set, respectively. This result is
consistent with the fact that for a slip in the shuffle set, only one covalent
bond must be broken compared to three in the glide set.\cite{Sho53PR} 

In the high temperature domain, the probability of  dislocation nucleation
tends to drop and plastic strains taking the form of amorphizations occur. As
temperature is raised, the strain at which some plasticity occur, decreases and
this process lasts until the thermal vibrations are sufficient to begin the
melting/amorphisation and the applied strains to small to initiate a
dislocation in the shuffle set. Here again, no dislocation is formed in the
glide set. Conversely to our simulations, at high temperature, the observed
dislocations are partial dislocations belonging to the glide set planes.  It is
commonly accepted that they  move more easily through the nucleation and
propagation of double kinks thanks to thermal vibrations. \cite{Hir82WIL,
Bul95PMA, Bul01PMA, Mit03PM} However, the size of the simulation cell along the
dislocation line used here, 4 a/2$<110>$, is too small to allow the formation
of a kink pair.  Consequently, only two plastic events are possible in the
simulation: the nucleation of  an infinite straight $60^\circ$ dislocation in
the shuffle set planes or amorphisation/melting, depending on the temperature.

Experimentally, the observations done in both low and high temperature domains
reveal a slip mode transition depending on the temperature. At low temperature
dislocations seem to glide in the shuffle set planes and at high temperature in
the glide set planes.\cite{Due96PML, Rab01MSE, Rab01SM2} Whatever the
temperature, our simulations have shown that the nucleation of straight
dislocation in the glide set plane is not allowed due to geometric reasons. The
only type of dislocation, the $60^\circ$, is nucleated in the shuffle set
planes. Moreover we have observed that high temperatures prevent the
dislocation formation in this set. Thus our results are not in disagreement
with the experimental facts, but complementaries calculations in 3 dimensions,
in order to allow the kink propagation, are necessary to confirm the slip mode
transition.

\subsection{Character of the dislocation nucleated.} 

In order to understand the kind of dislocation formed, we tried to establish
the main criteria that govern this choice. Usually, when a crystal is stressed,
the slip system with the largest resolved shear stress along the Burgers vector
{\bf b} is favored. In our case, the resolved shear stress on each dislocation,
proportional to the Schmid factor, is directly related to the stress
orientation $\alpha$.  In the range of temperature where  the glide events are
frequently observed for the SW potential, in most cases, the plastic events are
consistent with the results predicted by the Schmid factors. For example,  on
the Figure~\ref{slip_sys}b, the $60^\circ$ dislocation is favored for a stress
orientation $\alpha=22.5^\circ$ in traction and in compression, what is
obtained in our simulations. One sees from the same figure, that a compressive
stress disorientated by $36^\circ$ favors the $30^\circ$ partial, that is a
strain in the twinning sense.  Since dislocations of the glide set are not
activated, as explained above, the system finds another slip system to relax
the applied stress. In these conditions of twinning, two dislocations are
possible the $60^\circ$ and the screw. In our simulations, the dislocation
nucleated is the $60^\circ$, i.e. the one with the  largest Schmid  factor
(Fig.~\ref{Schmid}).

However several cases cannot be explained on the basis of the Schmid factor
only,  the character of the dislocation must also be taken into account. For
example, under a non disorientated tensile stress favoring the twinning stress
along the  {\bf b}$_{90^\circ}$ partial (Fig.~\ref{slip_sys}b), a crystal crack
is produced without glide events. Following the Schmid factor analysis, two
$60^\circ$ on both sides of the partial dislocation could then be nucleated.
But the resolved shear stress along both symmetric $<110>$ directions
(Fig.~\ref{Schmid}) are equal, what may prevent the choice of one slip system.
To check this, a calculation with a stress orientation at $10^\circ$ that
breaks the symmetry of the problem, has formed  a $60^\circ$ dislocation in
agreement with the Schmid factor. 

Compare this case to that where the stress orientation is disorientated by
$45^\circ$ (Fig.~\ref{Schmid} curves 3 and 5). Although the resolved shear
stress is the same on the screw and the $60^\circ$, the latter is nucleated, in
compression. It is worth remarking that  the two types of dislocations have
different mobility properties, cf. for instance  the Peierls stresses. The
calculations performed with the SW potential have shown that the Peierls stress
on  the $60^\circ$ dislocation is smaller than on the screw. \cite{Ren95PRB} To
relax the applied stress, the nucleation of a perfect $60^\circ$ dislocation is
then favored. 

The other discrepancies between the Schmid factor analysis and the simulation
results are mainly due to the unphysical defect created by the SW potential,
the micro-twins, which pollute largely the results in both traction and
compression, when the applied stress acts in the anti-twinning sense. For
example in compression at 0K, the micro-twin formation disappears as the stress
orientation $\alpha$ increases. Hence the resolved shear stress along the
anti-twinning direction must be as small as possible to avoid this kind of
defects. 

The analysis of the plastic strains as a function of the stress orientation
shows that the character of the dislocations nucleated from surface steps can
be mainly predicted by examining the Schmid factor and the Peierls stress. 
Other factor may play a role, though. For example, the crystal symmetry may
prevent the choice of one particular slip system leading to fracture.

\section{Conclusion}

We have investigated the nucleation of dislocations from linear surface defects
such as steps, when the system is submitted to a uniaxial stress. Although the
elastic limits remain relatively close to the theoretical strength, it appears
that the surface steps weaken the atomic structure and help the formation of
glide events like dislocations. The glide evens are nucleated and propagated in
the planes of the shuffle set. No straight dislocation is formed in the glide
set plane. The geometry of the simulation cell used here which prevents the
formation of kink pairs, does not allow for the expected formation of partial 
dislocations in the shuffle set at high temperature.

In addition, we have remarked that the high temperature decreases the
probability of nucleating perfect dislocation in the shuffle set plane.
Melting/amorphisation of  silicon occurs before reaching the required  shear
stress to initiate the dislocation.  These results seem consistent with the
assumption that at low temperature the dislocations glide in the planes of the
shuffle set, based on the observation of non dissociated dislocations  in
silicon samples deformed at low temperature in conditions preventing failure.
\cite{Rab01MSE,Rab01SM2} Supplementary studies are planned to check the
nucleation of dislocation loops in the glide set planes with high temperature.

The role of the stress orientation on the nucleated defects has been studied
from the calculations performed with the SW potential. Although the results are
slightly biased by the somewhat unphysical defect produced by the SW potential
when the stress acts in the antitwinning direction, it emerges that the type of
dislocation nucleated is chosen by the resolved shear stress and the Peierls
stress. 

Concerning the empirical potentials, it has not been possible to nucleate any
dislocations in the simulations performed with the Tersoff potential and EDIP
at 0K. The Tersoff potential has too high energy barriers preventing the bond
breaking required to nucleate a dislocation at low temperatures. While EDIP
presents a shear instability in the shuffle set planes. With the Tersoff
potential, the overcoming of the energy barriers leading to the dislocations
nucleation has become possible at high temperature. By extrapolation, EDIP is
probably able to nucleate dislocations thanks to the thermal vibrations. To
summarize, although the different results are potential-dependent, only the
simulations performed with the SW potential can be taken into account at 0 K as
demonstrated in our previous study on bulk system. Actually, we are trying a
similar calculation with ab initio methods on a relatively large system. A
calculation with a small system of about 200 atoms has already produced the
nucleation of a $60^\circ$ perfect dislocation in the shuffle set.




\section*{Table captions}

\begin{table*}[h] 
\caption{Summary of plastic events obtained with the SW potential, for several
stress orientations at 0K and with temperature. All the glide events are
localized in the shuffle set planes. The elastic limits are given at 0K by the
uniaxial stresses. Note that, the strains along the stress direction are
obtained by the linear elasticity.} \label{SW_results} 
\end{table*}


\begin{table*}[h] 
\begin{ruledtabular}
\begin{tabular}{ccddcc} 
$\alpha$ & \multicolumn{2}{c}{Stress (GPa)}& \multicolumn{1}{c}{strain (\%)}& Results T = 0K & T $\alt$ 900K\\
\hline 
$0^\circ$ & Trac &31.5&22.9& fracture & fracture \\
	  & Comp &-10.5&-7.6& micro-twin & micro-twin\\
\hline 
$10^\circ$ & Trac &25.5&19.1& fracture & perfect $60^\circ$ then fracture \\
	   & Comp &-10.5&-7.9& micro-twin & micro-twin \\
\hline 
$22.5^\circ$ & Trac& 22.5&18.7&perfect $60^\circ$  &perfect $60^\circ$ \\
	   & Comp &-12.0&-10.0& plastic deformations in \{111\} planes & perfect $60^\circ$ \\
\hline 
$36^\circ$ & Trac &21.0&19.2& micro-twins + sometime $60^\circ$ and screw & micro-twins + large strained zone \\
	   & Comp &-13.5&-12.4& perfect $60^\circ$ & perfect $60^\circ$ \\
\hline 
$45^\circ$ & Trac &21.0&19.7& micro-twins & micro-twins \\
	   & Comp &-15.0&-14.0& perfect $60^\circ$ & perfect $60^\circ$ \\
\end{tabular} 
\end{ruledtabular}
\end{table*}

\vfill\centerline{\bf Table~\ref{SW_results}}

\newpage

\section*{Figure captions}

\begin{figure*}[h] 
\caption{Diamond-like structure projected along $[0\bar{1}1]$ (a) and along
$[111]$ (b). All the possible slip directions following the Burgers vectors of
the $60^\circ$, $90^\circ$, $30^\circ$ and screw dislocations are considered. 
For each dislocation, the best stress orientation giving the maximum resolved
shear stress is noted.}\label{slip_sys}  
\end{figure*}

\begin{figure}[h] 
\caption{Calculation cell with a $D_B$ step non re-bonded. $\bm{\sigma}$ is the
applied uniaxial stress and $\alpha$ the angle between the $[011]$ normal step 
and the stress direction.}\label{cell} 
\end{figure}

\begin{figure*}[h] 
\caption{The Schmid factors versus the stress orientation $\alpha$ are drawn
for five slip directions; two along the Burgers vectors of the $60^\circ$,
$<10\bar{1}>$ (1) and $<1\bar{1}0>$ (3), two along the Burgers vector of the
partials, $<2\bar{11}>$  for the $90^\circ$ (2) and $<1\bar{2}1>$ for the
$30^\circ$ (4) and the last one along {\bf b}$_{screw}$, $<0\bar{1}1>$ (5), 
parallel to the step line. (the Schmid factor being proportional to the
resolved shear stress). }\label{Schmid}  
\end{figure*}

\begin{figure}[h]  
\caption{Nucleation of a perfect $60^\circ$ dislocation from the surface step,
in a plane of the shuffle set with the SW potential. The tensile strain is
about of 18.7\% and disorientated of $22.5^\circ$.}\label{sw_22.5_tract}  
\end{figure}

\begin{figure}[h]  
\caption{Nucleation of a perfect $60^\circ$ dislocation in the shuffle set
plane thanks to the temperature. The compressive stress is disorientated of
$22.5^\circ$. (a) SW: the $D_B$ step disappears for a strain of -7.5\% at 900K,
(b) Tersoff: one atomic layer disappears for a strain of -11.0\% at
1200K.}\label{disloc_temp}  
\end{figure}

\begin{figure}[h]  
\caption{Snapshot of silicon structure close to the elastic limit with a stress
disorientated of $22.5^\circ$. (a) Tersoff with a strain of 46.7\%, (b) Tersoff
with a strain of -38.5\%, (c) EDIP with a strain of 34.5\%, (d) EDIP with a
strain of -8.9\%.}\label{ters_edip}  
\end{figure}

\newpage

\begin{figure*}[h] 
\includegraphics[width=16cm]{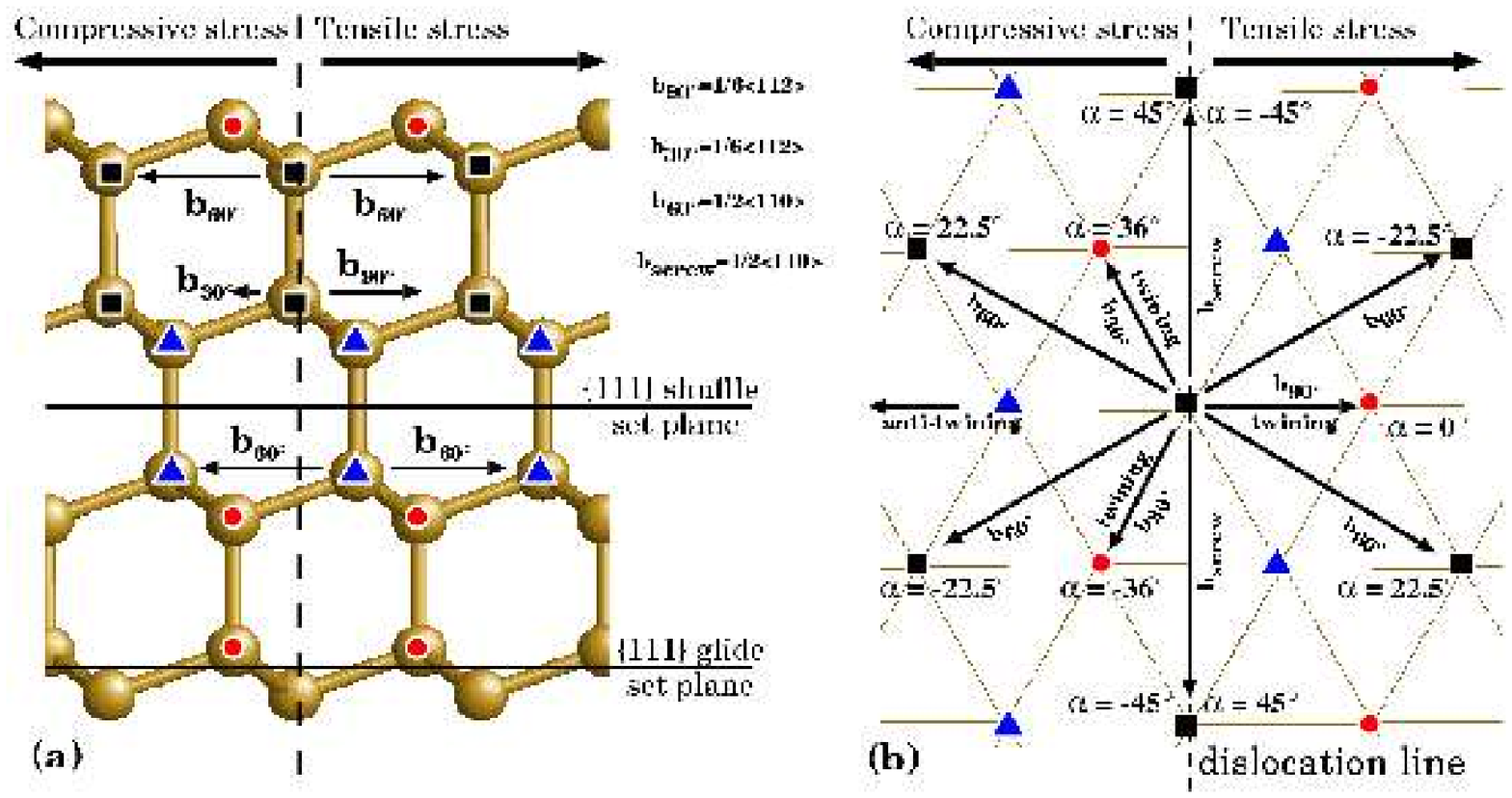} 
\vfill\centerline{\bf Figure~\ref{slip_sys}}
\end{figure*} 

\begin{figure}[h] 
\includegraphics[width=8.6cm]{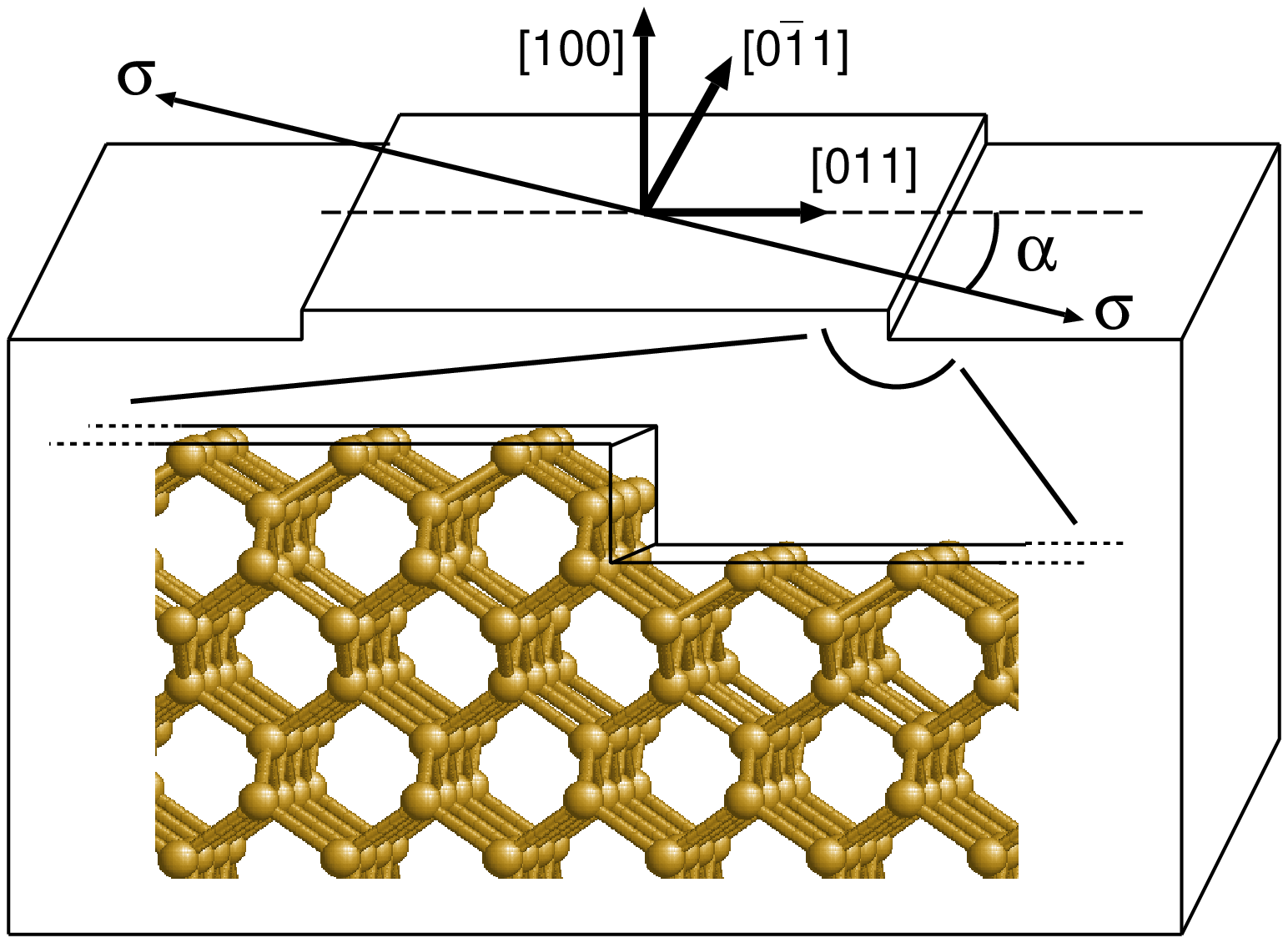} 
\vfill\centerline{\bf Figure~\ref{cell}}
\end{figure}

\begin{figure*}[h] 
\includegraphics[width=16cm]{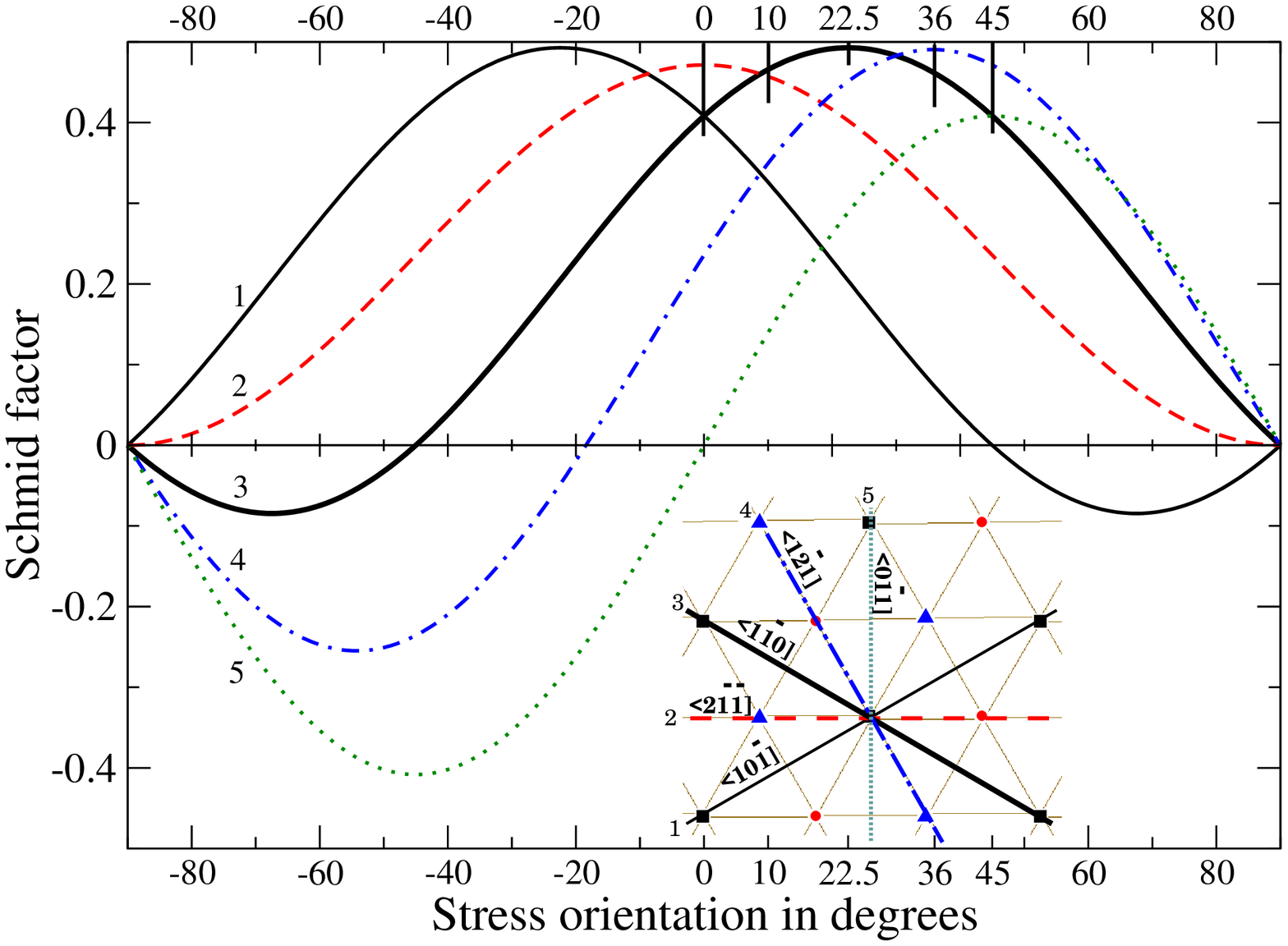} 
\vfill\centerline{\bf Figure~\ref{Schmid}}
\end{figure*} 

\begin{figure}[h] 
\includegraphics[width=8.6cm]{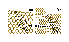} 
\vfill\centerline{\bf Figure~\ref{disloc_temp}}
\end{figure}

\begin{figure}[h] 
\includegraphics[width=8.6cm]{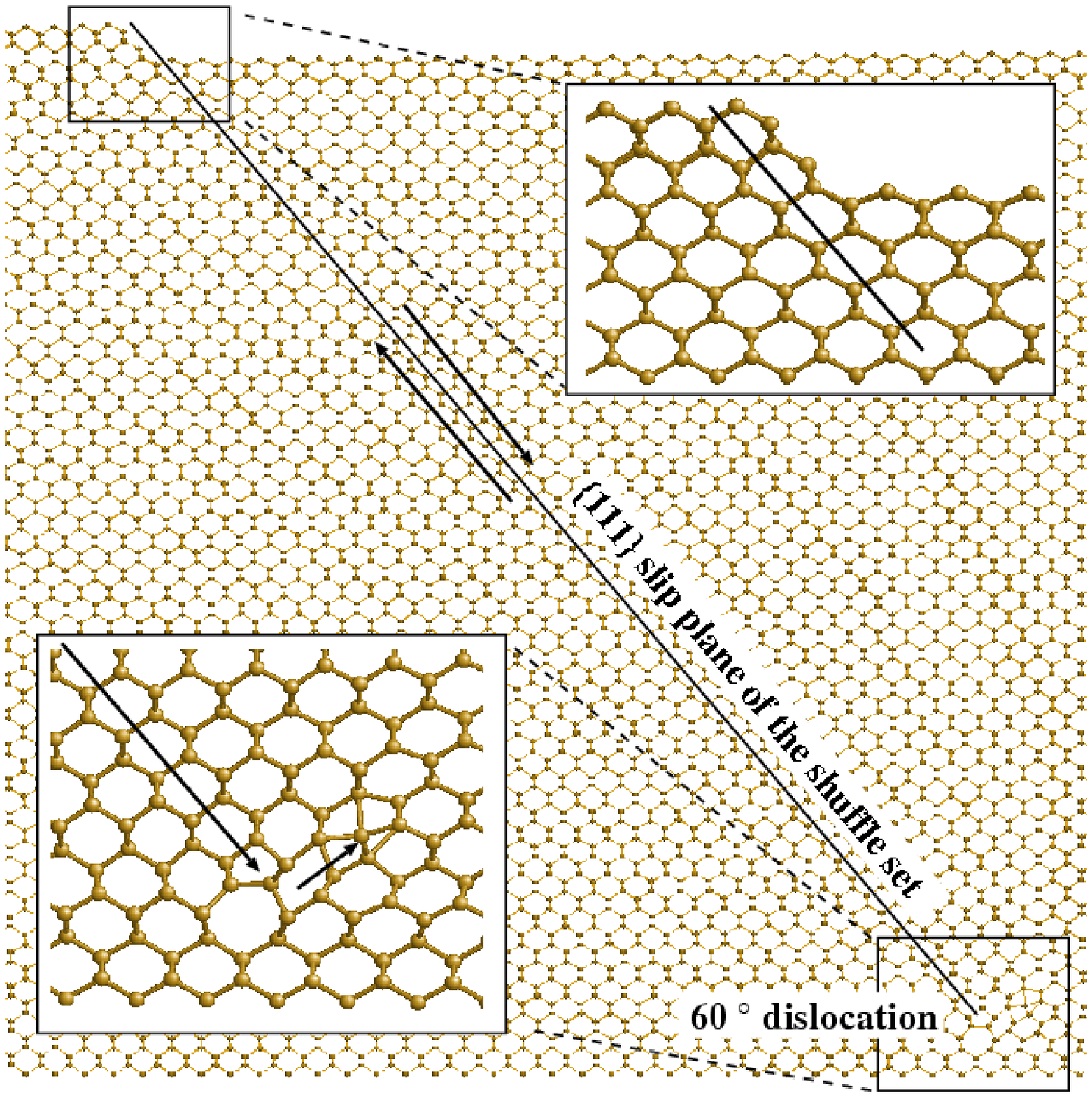} 
\vfill\centerline{\bf Figure~\ref{sw_22.5_tract}}
\end{figure}

\begin{figure}[h] 
\includegraphics[width=8.6cm]{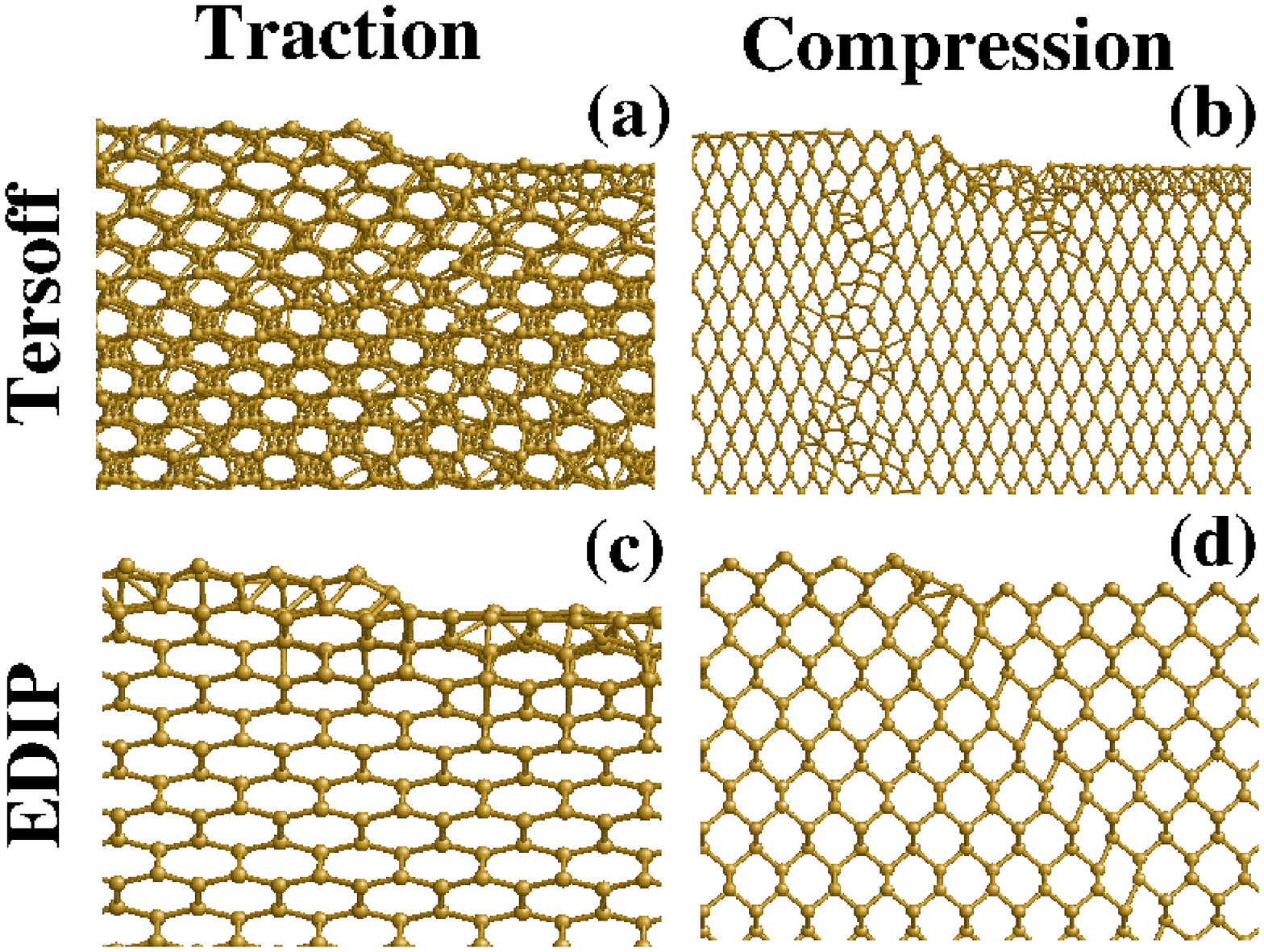} 
\vfill\centerline{\bf Figure~\ref{ters_edip}}
\end{figure}

\end{document}